\documentclass[12pt]{iopart}
\usepackage{epsfig}
\newcommand{\be}{\begin{equation}}
\newcommand{\ee}{\end{equation}}
\newcommand{\hmp}{Mpc\,h^{-1}}

\begin{document}
\title{The Holtsmark distribution of forces 
and its role in gravitational clustering}
\author{L. Pietronero \dag \ddag, M. Bottaccio \dag \ddag,
R. Mohayaee \ddag
and M. Montuori \ddag}
\address{\dag\ Dipartimento di Fisica, Universit\'a di Roma ``La Sapienza'',
Pl.Aldo Moro 2, 00185 Roma, Italia }
\address{\ddag\ INFM, Section Roma 1, Pl.Aldo Moro 2, 00185, Roma, Italia}
\ead{mb@pil.phys.uniroma1.it}

\begin{abstract}
The evolution and the statistical properties of an infinite gravitating
system represent an  interesting and widely 
investigated subject of research. 
In cosmology, the standard approach is based on 
equations of hydrodynamics.
In the present paper, we analyze the problem from
a different perspective, which is usually neglected.
We focus our attention on the fact that at small scale
the distribution is point-like, or granular, and not fluid-like.
The basic result is that the discrete nature of the
system is a fundamental ingredient to understand its evolution.
The initial configuration is a Poisson distribution in which the
distribution of forces is governed by the Holtsmark function.
Computer simulations show that the structure formation
corresponds to the shift of the granularity from
small to large scales. We also present a simple cellular automaton
model that reproduces this phenomenon.
\end{abstract}

\maketitle

\section{Introduction}

The basic mechanism for the emergence of coherent structures in 
gravitational N-body simulations still presents important open problems,
 in spite of 
extensive studies in the past few decades.
Indeed, the   problem is particularly complex in various respects:
\begin{itemize}
\item {\em conceptual}. The gravitational
 potential  is always attractive, it has infinite range
and diverges in the origin. These features do not
allow a straightforward application of the standard statistical mechanics.
For example, the standard way to perform the thermodynamic limit 
(the limit in which the size of 
system and the number of particles go to infinity, while the density is kept 
constant) is to study the intensive features of a large, but finite system
versus the number of particles. For systems with short range interactions,
these properties are usually finite in the limit.
If we apply the same procedure on gravitating systems, however,
the thermodynamic limit of these quantities is infinite;
\item {\em computational}. As a consequence of the
 infinite range of the interaction, each particle interacts
with any other in the system. Therefore  the number
of computations to evaluate the force on each particle
is $O(N^2)$ in a system with $N$ particles.
Furthermore, the treatment
of close encounters requires a high accuracy, because
of  the divergence in the origin. Both these elements
 result in  particularly heavy simulations.    
\item {\em physical aspects}. In particular, it is 
not clear whether the evolution of a self gravitating
system depends crucially on initial conditions, or, 
on the contrary, it shows some sort of self-organization
in the dynamics.   
\end{itemize}

One of the strongest motivations for studying such systems and
their dynamics 
is to understand the observed distribution of galaxies in the 
universe

Galaxies 
are  actually arranged in 
complex structures,  such as filaments and  walls
separated by large voids. 
The  statistical
features of such large-scale structures 
and the understanding of their formation
is one of the main topics of the modern cosmological research.
The starting point can be to
describe correctly the statistical properties
of the spatial distribution of the galaxies.
In  the past years this issue 
stimulated a large debate, concerning the suitable statistical
tools to apply in such study. 
Roughly speaking we can identify two main points
of view: 
the first, that is most popular, claims that 
the galaxy distribution exhibits fractal 
correlations with dimension $D \approx 1.3$ at  small 
scale (i.e. up to $\sim 10 \div 20  \hmp$), while at 
larger scales there is overwhelming evidences in 
favor of homogeneity \cite{davis97}, \cite{wu99}
This conclusion is based on the analysis  of the 
two point correlation function $\xi(r) $, first 
introduced in the analysis of galaxy catalogs
 by Totsuji $\&$ 
Kihara (1969) \cite{Toshui}  and 
then widely applied and studied by many others as 
P.J.E.Peebles~\cite{Peebles}.

A completely different interpretation 
of galaxy correlations has been suggested 
by Pietronero and collaborators 
\cite{coleman:pietronero:92}, \cite{sylos:montuori:pietronero:98}.
By analyzing the different galaxy samples 
with the statistical tools suitable 
for the  characterization of irregular 
(and regular) distributions, it has been 
found that galaxy 
and cluster  distributions exhibit 
very well defined scale invariant 
properties with fractal dimension 
$D \approx 2 $ up to the limits 
of the analyzed samples ($R_s \approx 50\div 100 \hmp$).

In conclusion 
there is a general agreement that 
the inhomogeneities seen in galaxy catalogs 
correspond to a correlated fractal distribution 
up to $10 \div 20 \hmp$, but both the value of the 
associated fractal dimension and 
the properties at larger scales 
remain controversial.

In any case, fractal correlations appear 
to not be included easily in the standard 
framework of gravitational 
structure formation, 
which is usually based on a fluid 
description of the matter density field. 
In these models, the dynamics 
is described by Euler, continuity and Poisson equations
for compressible fluids and the outcome 
strongly depends on the initial conditions.


Fractals are indeed highly non analytic structures, 
which are usually
the result of complex and highly non-linear dynamics.
Accordingly, we would like to approach the general issue of the
formation of structures from a point of view which is closer 
to the modern approach of critical phenomena.
In other words, we want to study a non perturbative,
fully dynamical approach of gravitational clustering.

Considering the observational results of a fractal structure 
in galaxy distribution we investigate if a system of point masses
evolves in a way which is not predicted by the fluid approach 
because of its discrete nature.
Actually, we would like to put in evidence the
role of non analyticity  in the formation of structures in
gravitating systems. 
We want here to tackle a specific point in the problem 
from a conceptual point of
view, retaining only the essential  elements of the problem.

Therefore we  analyze the evolution of
an ideal gravitating system with very simple initial conditions.
We do not consider additional elements which are usually  present
in cosmological N-body simulations 
such as  cosmological expansion, 

We study a system with the following characteristics:
\begin{itemize}
\item $N$ particles with the same mass
\item random initial positions
\item cubic volume with 
periodic boundary conditions,
{\em both for the motion of particles and for the total force
 acting on each particle}; in this way there is no
special point in the system and forces from far away are taken into account,
so the long-range behavior of the interaction is retained;
\item simple initial velocity distributions. Typically we use
zero initial velocities.
\end{itemize}
 
Periodic boundary conditions allow to some extent
 to mimic an infinite system, and prevent the system from
having an a-priori preferred point. It has to be stressed, 
however, that, despite
being infinite, the system has in fact a finite number of degrees of freedom.
 In order to understand the behavior of the true infinite systems, we have
performed  N-body simulations with the same number density but with different
number of particles. This is in the spirit of the {\em thermodynamic
limit}, and should allow us to identify finite
size effects. 


A long-debated point is that the potential in a
point in an infinite non expanding system
is a divergent quantity, as it can be verified with 
a naive computation. Actually, for an infinite,
homogeneous distribution of matter, the potential at a point would 
diverge as $R^2$ for $R\rightarrow \infty$:
Correspondingly, the ordinary Poisson equation gives unacceptable results.
Such difficulty has often been  raised to argue for the
unphysical nature of such models.
A formal solution to this problem was proposed by Jeans very long ago
\cite{Jeans}. He suggested to replace Poisson equation with
\be
\label{eq:Poisson}
\nabla^2 \phi(\vec r)=4 \pi G (\rho(\vec r) -\rho_0)
\ee
where $\phi$ is the potential, $G$ is the gravitational constant,
 $\rho(\vec r)$
is the local density  and $\rho_0$ is the average density of 
the system.
In such a way the potential is well defined and it is not divergent.
This trick ({\em Jeans' swindle}) has been largely criticized, anyway,
because the addition of the extra term in the right hand side
 of equation (\ref{eq:Poisson})
 looks rather artificial.

This is not a problem in the presence 
of cosmological  expansion, since
 the expansion effectively removes the divergence (e.g.~\cite{Peebles}),
and an equation analogue to (\ref{eq:Poisson}) can be found.

Here, we want to take a very pragmatic attitude.
What we really need in our simulations  and in the real systems too,
is the force acting on a particle of the system. 
Due to large scale statistical symmetry in our
systems, this quantity is well defined, since large scale contributions
average out.
It is interesting to recall that the convergence of the force is not
assured for any system.  It can be
diverging for some distributions of matter.
Interesting examples  are fractals with
fractal dimension
$D>2$ \cite{Gabrielli}; in this case the intrinsic anisotropy of the system
is such that long range contributions do not balance.
 For gravitating fractals with $D\,<\,2$, the force is
convergent, since even if the distribution of points
is anisotropic, the infinite number of contributions to the force by distant
particles sum up to a  finite value, because of the fast
decreasing density of objects.

For the systems we are interested in, the force is in fact
conditionally convergent. In other words, the force converges 
considering spherical shells centered on the particle of the system.
On the contrary, if one  computes the forces  by summing the contributions in 
different ways (i.e. considering first all those coming from particles on 
one side, then all those coming from particles on the other side),
we  obtain
a diverging or undetermined result.

 
If we were to need the potential,
we may {\em define} it as the function which is well-defined (i.e.,
non-diverging everywhere) and whose gradient gives the right expression
for the force. One can verify that the potential in equation
(\ref{eq:Poisson})
 satisfies these requirements.

We used N-body simulations to study the evolution of such system, 
and analyzed its statistical properties.

\section{The numerical code}
We use two distinct numerical algorithms for our simulations;
they are  described in 
\cite{Tesi},\cite{Gadget}. Here it is enough to recall that the
code we use solves coupled equations of motion by discretising time.
In this way we are able to follow the motion of each individual particle.
 
Basically, the code:
\begin{enumerate}
\item evaluates the force acting on each particle. We used a 
numerical method called
{\em tree code} is used which speeds up the computations by
performing appropriate approximations. Periodical 
boundary conditions are included
by means of Ewald formula~\cite{Ewald}.
 This method allows to include large scale effects, due
to infinite replication of large overdensities by a truncated
sum in Fourier space,
while reproducing the relevant particle-particle interactions by a
truncated sum in real space.
\item It advances the particles for a discrete time step using
such forces by a second order numerical integrator ({\em leap frog
or Verlet scheme}). To save computation time, we use a different
time step for each particle. A particle whose  dynamical state is
changing fast has a very small time step, while a particle which is
slow or slowly accelerating has a proportionally larger step.
\item It repeats the previous points.
\end{enumerate}
We introduce a smoothing in the potential at small scales 
to avoid time steps too small (which would result in a very long simulation).
This means we modify the potential at very small scales, such that
the divergence in the origin is removed. Roughly speaking,
we substitute the $-1/r$ potential on scales smaller than the
smoothing length $\epsilon$
 with a monotonically increasing function  matching
$-1/r$ at $\epsilon$ and with zero derivative
in the origin.

This procedure is quite customary in astrophysical simulations 
(e.g.~\cite{Hern89}),
but one has to take care of how the smoothing affects the dynamics.
In some cases the smoothing can also be useful to model particular
situations. Typical examples are cosmological simulations
for the formation and evolution of galaxies. Such processes involve
a huge number of elementary particles,
of the order of $10^{70}$. 
To simulate such system one usually represents it
by  some new particles, chosen such that
 each corresponds a large chunk of elementary particles. The simulations are
then run using such particles.  A large smoothing here prevents
them to undergo the hard collisions typical of the two body
encounters, which would be unphysical in such situation. Since the
dynamics of the system is forced to be much smoother than in our case,
such simulations don't require a very careful time integration, i.e.
very small time steps; therefore very large simulations can be run in a
reasonable time. It is not clear, however, how faithfully such
simulations can actually describe the behaviour
of the original system \cite{sylos,melott}.
 
In our case, though, since we would like to study a pure gravitational
system, the smoothing should have very little or no influence
on the statistical properties we look for.
We have verified this by running  $16^3$ particles with
smoothing lengths smaller than the one we use, and the results were
in agreement.

\section{Discussion of the simulations}
The statistical tool we use is the standard one
for the study of the correlation properties of a generic 
point distribution, i.e. the {\it average conditional galaxy density}
 $\Gamma^*(r)$ versus 
the scale $r$,
\begin{equation}
\label{gamma}
\Gamma^*(r) = \frac{<N(<r)>}{4/3 \pi r^3 }
\end{equation}
$N(<r)$ is the number of galaxies contained in a sphere of 
radius $r$ centered on a galaxy of the sample and  
$<N(r)>$ is the average of $N(<r)$  
computed in all the spheres centered on every 
galaxy of the sample.

\begin{figure}
\centerline{
        \psfig{file=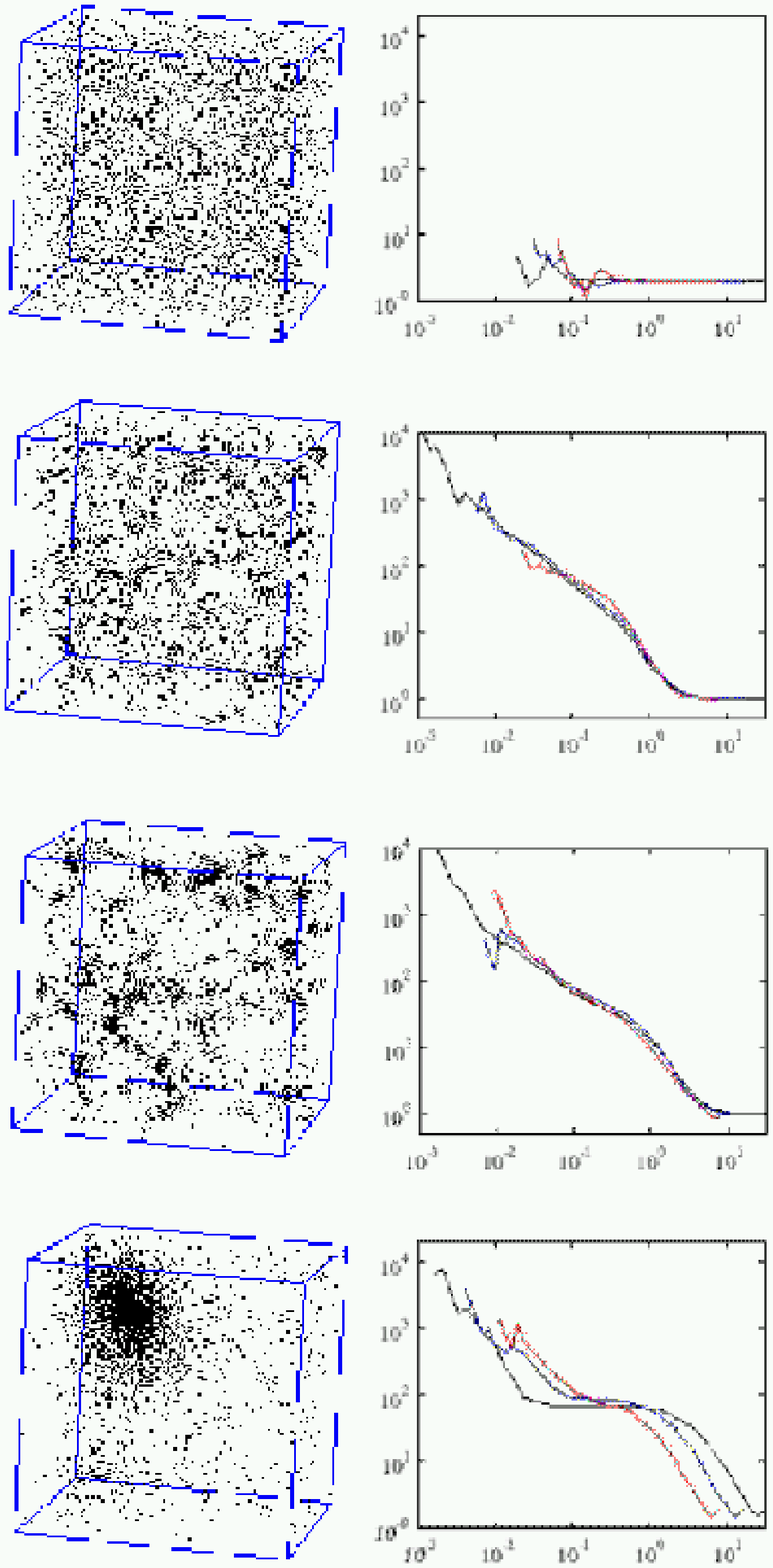,width=12cm}
}
\caption{Evolution of particle positions for a simulation with $32^3$
particles (left column) and the corresponding measure of $\Gamma^*$ 
(right column). 
The three lines  correspond to the $\Gamma^*$  for
simulations with $N=8^3$, $N=16^3$, $N=32^3$ particles. 
}
\label{fig:evol}
\end{figure}

\begin{figure}
\centerline{
        \psfig{file=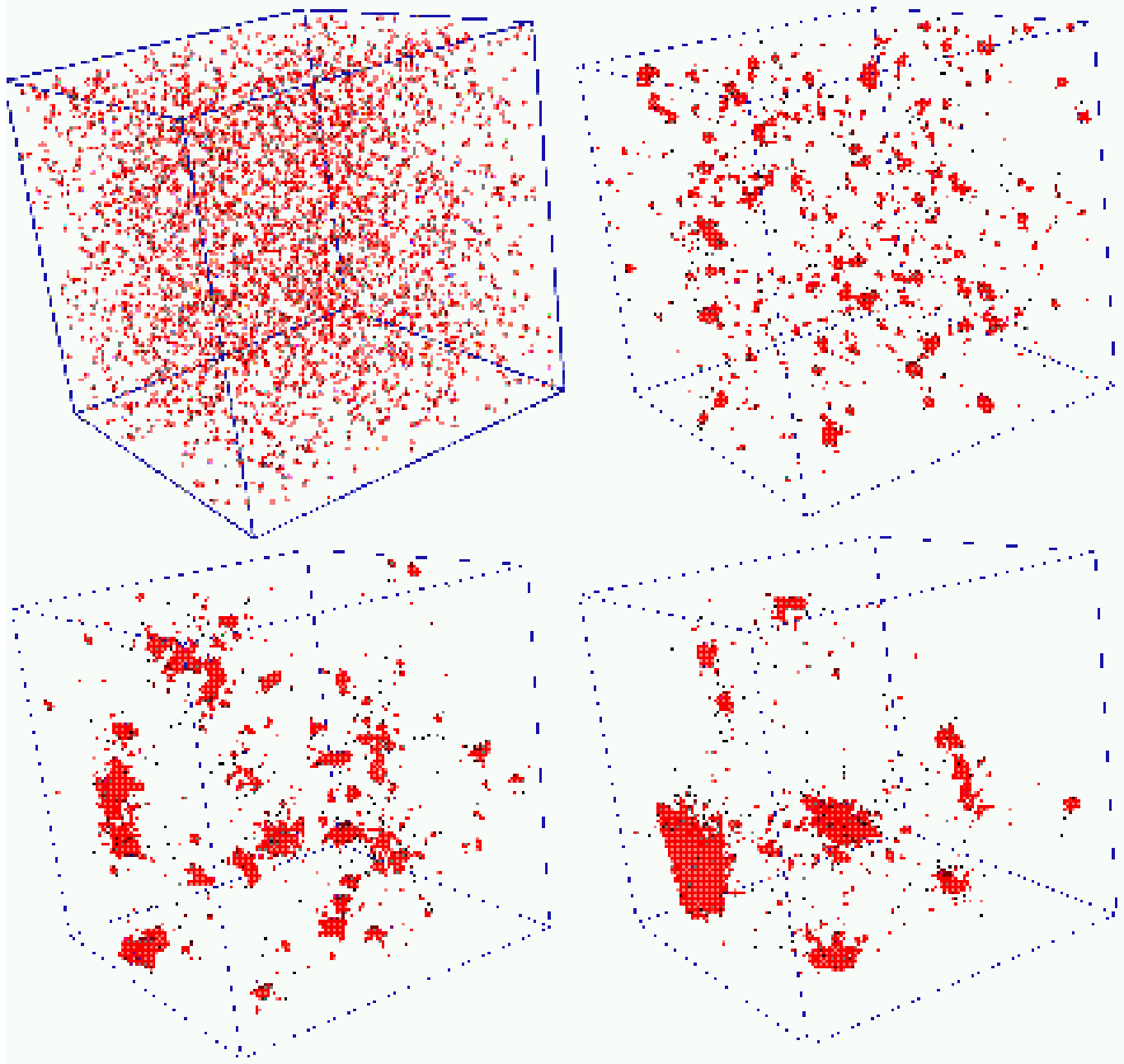,width=11cm}
}
\caption{Evolution of particle positions for a simulation with $32^3$
particles. 
}
\label{fig:evol_corr}
\end{figure}
 
We run a series of simulations with $N=8^3$, $N = 16^3$, 
$N = 32^3$. 
The evolution of the positions of the particles for the 
$N=32^3$ simulation is shown in the left column of figure \ref{fig:evol}.
At the initial time the system is set with a homogeneous density and
no velocities. After a while, clusters begin to form at small scales,
while at large scales the distribution is still homogeneous.
As the evolution goes on, clustering extends to larger and larger scales,
up to the size of the simulation box. In the final stage of the system almost
all the particles belong to 
a single cluster. This state is stationary.   
 
In figure \ref{fig:evol} we show the evolution of particle positions 
(left column) together
with the corresponding measures of spatial correlations, performed
with $\Gamma^*$ (right column). 
The three lines  correspond to the results for the aforementioned  
three simulations.  
An important consideration is that the correlation properties are the  
same for systems with different number of particles. 
This is true for the intermediate states of the evolution, of 
which we show two snapshots at different times in the central  
rows of fig.\ref{fig:evol}. 
On the contrary, this is not true  for the last row, 
which corresponds to the final, stationary state of the systems. 
 
We can interpret such results by saying that the intermediate stages 
of the evolution show a well defined {\em thermodynamic limit},  
that is, the correlation properties at a fixed time are the same 
as $N\rightarrow\infty$, keeping $N/V=const$. 
This property can be understood by noting that the effective range of the 
gravitational interaction is finite.

This seems to be in  contradiction with
the long range nature of the interaction. 
However, the reason  lies in the  
long range isotropy of the initial conditions. 


\section{The Holtsmark distribution}
The distribution of the forces acting on a particle in
an infinite Poisson system can be derived exactly~\cite{holts},
\cite{chandrasekhar43}.
The starting point is to consider the probability 
of having a force $F$ on a particle due to 
all the $N$ particles in a sphere centered on it. 
Such probability can be expressed as:
\be
W_N({\bf F}) 
=
\int d^3 \,x_1 \ldots d^3\, x_N\, p({\bf x_1})\ldots p({\bf x_N})
\delta({\bf F} -\sum_{i=1}^N {\bf F(x_i)}) 
\label{holts:1}
\ee
where $p({\bf x_i})=3/(4\pi\,R^3))$  is the probability to
have particle $i$ in ${\bf x_i}$.
In the equation (\ref{holts:1}), it has been
possible to write the $N$ point probability as 
the product of $N$ independent one-point 
probabilities, because the point distribution is
Poissonian.
The dirac $\delta$ can be now expressed in its Fourier 
representation.
This allows to reduce the previous integral to:
\be
W_N({\bf F})
=
\left(\frac{1}{2\pi}\right)^3
\int exp (- i {\bf k}\cdot{\bf F}) A({\bf k}) \, d^3k
\ee
where:
\be
A({\bf k})
=
\left[
\int exp(i{\bf k}\cdot{\bf F(x)}) p({\bf x}) d^3 x
\right]^N
\ee
and the last integral extends over the sphere.
Expressing $N$ as $N= 4\pi R^3 n/3$, where $n$ is the average density 
in the system, and taking the limit $R\rightarrow \infty$ and $n= const$
 $A({\bf k})$ becomes:
\be
A({\bf k})
= 
\exp(-nC({\bf k}))
\ee
where 
\be
C({\bf k})=
\int_{|{\bf r}|}^{+\infty}d^3r
(1-\exp({i {\bf k}\cdot{\bf F(r)} })
\ee
After some algebra we eventually get  {\em Holtsmark distribution}:
\be
W(F) = \frac{2}{ \pi F}
 \int_0^{\infty} dx\; \exp[ -(x/\beta)^{\frac{3}{2}}]\;x\;\sin(x) \;
\label{h1}
\ee

where
 $ \beta = F/F_o$ is an adimensional force
 and
$F_o=\left[ 2\pi (4/15)^{2/3}\right] Gm/\lambda^2$. In the latter expression,
$\lambda$ is the average inter-particle separation $=(V/N)^{1/3}$.
In the limit $F\rightarrow \infty$, the equation (\ref{h1})
can be written as:
\be
W(F)
=
\frac{2}{ \pi F}
Im \int_0^{\infty} x e^{ix}\left[ 1- (x/\beta)^{\frac{3}{2}} +\ldots\right]dx
\label{h2}
\ee
Performing a suitable change of variables and retaining only the 
dominant terms, we get:
\be
\lim_{F\to +\infty} W(F)
=
\frac{15}{8}\left(\frac{2}{\pi} \right)^{1/2}\frac{1}{F_0}\beta^{-5/2}
=
2\pi n\left( Gm \right)^{3/2} F^{-5/2}
\label{h3}
\ee

On the hand, it is simple to compute the distribution of the forces
$W_{nn}(F)$
on a particle due to its nearest neighbour in a Poisson distribution.
Actually, in this case:
\be
W_{nn}(F)dF = \omega(r)dr
\label{h4}
\ee
where $\omega(r)= 4\pi r^2 \, exp( -4\pi r^3 n/3)$ is the 
probability to find the nearest neighbour of a particle at 
a distance $r$ from it.
From the previous equation we get:
\be 
W_{nn}(F) = 4\pi r^2n \cdot \exp( -4\pi r^3 n/3)\left(\frac{dF}{d r}\right)^{-1}
\label{h5}
\ee
which can be easily solved:
\be
W_{nn}(F) = 2\pi n\cdot(Gm)^ {3/2} F^{-5/2}
\exp\left[-4\pi n/3\cdot \left(Gm \right)^ {3/2}F^{-3/2}\right] 
\label{h6}
\ee
In the limit $F\rightarrow \infty$ (i.e. $r<< \lambda$) the solution is 
simple:
\be
W_{nn}(F) \approx 2\pi n\cdot(Gm)^ 
{3/2} F^{-5/2}
\label{h7}
\ee  
The two expressions (\ref{h3}) and (\ref{h7}) 
are actually identical. 
This implies that the largest 
forces on the particles of the system are those 
due to the nearest neighbours. Particles quite isolated with respect 
to any another in the system are instead submitted to a force which 
comes from the interaction with a large number of particles

\section{A qualitative model}

At the beginning, the first particles to move are 
those which experience the largest forces.
As we have seen, when the initial point distribution is
poissonian, such particles 
are those with a neighbour closer than the average. The 
dynamics
in this stage is then dominated 
by interactions on the scale of the
nearest neighbours. 
In this sense the gravitational interaction at this stage has a finite
effective range.
This two-body interaction breaks the initial isotropy on a scale slightly
larger than the mean inter-particle distance, while on large scales 
the system is still isotropic. 
Therefore, it is reasonable to argue that the dynamics we have described
will act on this new scale.

The relevant dynamics therefore takes place on the scale on which the  
system is anisotropic.  Once the particles have clustered on a scale, 
the isotropy is broken on a larger scales. Therefore, considerations  
similar to those developed for the scale of particles hold at subsequent 
times on larger scales, and it appears that the relevant forces are  
due to mass clumps present on such scales, while at very large scale the  
system is still isotropic. In this view long range interactions 
can be treated as interaction acting on a short range which increases with  
time. 
 
In such a sense, gravitational dynamics proceeds from small to large scales 
by a transfer of discreteness from small to large scales. 

When such scale is of the order of the simulation box,
the evolution is dominated by finite size effects. Indeed, the particles 
are grouped only in few clusters and the existence of periodic boundary
conditions, which prevent more particle to enter the simulation volume, 
forces the system toward a stationary state.
As a consequence, for a system with a larger number of particles
such finite size effects would take place at a later time, and therefore
on a larger scale, as can be seen
in the last row of figure~\ref{fig:evol}. Here, the rightmost line is related
to the simulation with $N=32^3$ particles, the leftmost line to the
simulation with $N=8^3$. 

Our description of the clustering  is based on the discrete nature
of the system. It is very interesting to notice that
in this model the evolution of a system with long range interaction
can be described in terms of  a series of short range processes,
which act on a progressively increasing scale.
We argue that this argument could allow to describe
the evolution of a true infinite self gravitating system at any time.

An important consideration which arises from this discussion is that 
this model of evolution {\em depends on the statistical properties of 
the initial conditions}. 
It applies to systems which are isotropic on large scales and anisotropic 
on small scales. 
On the contrary, for example, a typical initial condition used in cosmological 
simulations is a glass like distribution of particles, with large 
scale density fluctuations superimpose. 
In this case, the system is isotropic (regular) even on the smaller scales, 
so the dynamics is essentially driven by large scale fluctuations.

We take inspiration from renormalization to attempt to model the dynamics. 
The basic idea is that the relevant elements for the dynamics at a given time 
are the typical discrete objects formed at that time, i.e. clusters. 
We take the clusters to be structureless 
elements, and represent them with their total mass and their center of mass 
quantities. Fig.~\ref{fig:evol_corr} can help in imagining the system in this approximation. 
Now we can repeat the argument we made for particles to such elements, which 
we can take to be a sort of macro particles. Of course there is no exact analytic 
expression for the force distribution at this scale; it is reasonable however 
to assume the the nearest macro particle will be most important for the dynamics
of a given cluster. 
 
We can modelize the system at any time as made of almost independent pairs  
of macropartices, and assume that the relevant dynamics can be described in such a way. 
 
Visual inspection of the evolution of the system (fig.~\ref{fig:evol_corr})
suggests that clusters formed at a given time are the new discrete
elements for the iteration of the clustering process at a larger scale.

However, this model is based on  a strong assumption, 
i.e. that the clustering on a scales activates {\em after} 
the  clustering on smaller scales.
Of course this cannot be strictly true in general, since there are forces acting
between overdensities before clusters form on a scale. We assume nevertheless
that such forces are much smaller than those due to discreteness, so that the 
force due to it dominates the dynamics after some time. 

In \cite{simul} we have illustrated this model in  greater detail,
giving 
a qualitative estimate for the growth rate of the clustering process.


\section{A $1-d$ cellular automaton model}
As we have discussed in previous chapter,
the evolution of a random distribution of gravitationally 
interacting mass points
can be described by 
nearest-neighbours interactions taking place at increasingly larger
scales. 
This simplification is possible because the system is
reasonably isotropic on scales larger than
the scale of clustering.
To understand in greater detail the dynamics due
to a nearest neighbour interaction, we have investigated 
an extremely simple one-dimensional model.

We randomly distribute $10^5$ particles, of equal mass, on a 
line (of length $10^5$ units in our simulation), 
rather than on discrete lattice points as is the convention
in cellular automata,
and impose 
periodic boundary conditions. Particles move 
toward their nearest neighbours 
either by one unit at each time step or by half their separations 
from their nearest neighbours, whichever is shorter.  If
two particles are closer than a lower threshold, 
and in addition are {\it mutual} nearest neighbours, they 
coalesce at the mid-point of their separation, conserving mass.
If a particle is equidistant from its left and right
neighbours, which is an extremely rare event, it moves towards the right one.

Therefore this model  focuses precisely 
on the r\^ole of granularity and self-similarity 
and leads to a schematic picture for the gravitational clustering phenomenon. 
One of the main differences between  this model and
most existing aggregation models such as DLA, 
Smoluchoski or 1-dimensional Burgers, is that its dynamics depends
only on the positions and not on
the masses and initial velocities of the aggregates.

Our model suggests a simple interpretation for the non-analytic 
hierarchical clustering and can reproduce the self-similar 
features of the real gravitational N-body simulations 
for white noise initial conditions.

\begin{figure}[htb]
\centerline{
        \vspace{-1.8cm}
        \epsfxsize=0.4\textwidth
        \epsfbox{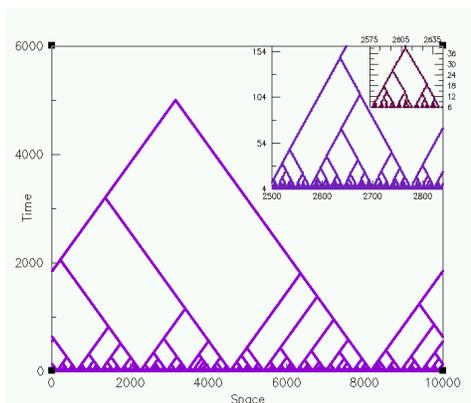}}
\vspace{1.2cm}
\caption
{The figure shows the trajectories of particles taken from 
a typical run of our simulation. The space-time tree
is reminiscent of river networks in two-dimensional space and 
exhibits the property of {\it topological scaling}, a concept
which is relevant for branched structures and originally arose 
as a means of analysing river networks.}
\label{tree}
\end{figure}

The trajectories in space-time of the aggregates are shown in fig.(\ref{tree}).
They show a tree-like topology.    
The tree structure of the aggregation process in space-time 
is a manifestation of topological self-similarity \cite{hasely00}, which is 
a property of many branched structures such as river-networks 
and bronchial trees \cite{rinaldo97} and can be quantified by 
various scaling exponents,  a most common of which  is  the Strahler 
index. For our model, this index has  a value of $2$, as   
in Cayley tree structures \cite{cmp00}. 

\begin{figure}[htb]
\centerline{
        \vspace{-0.8cm}
        \epsfxsize=0.58\textwidth
        \epsfbox{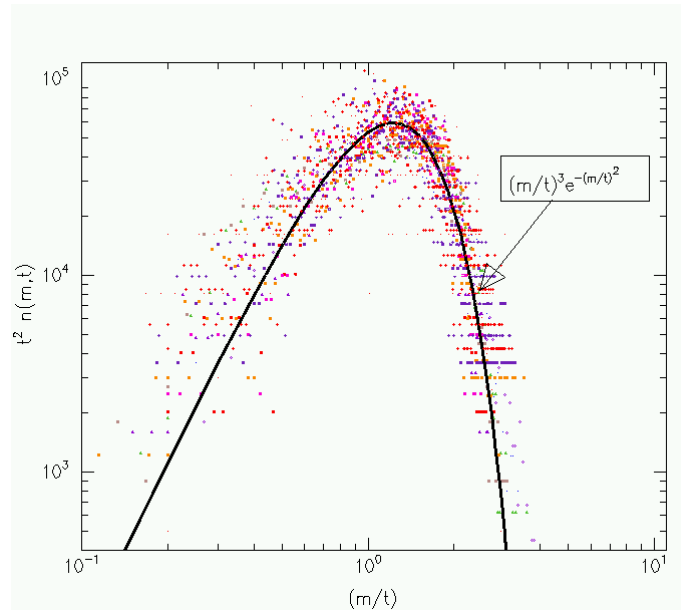}}
\vspace{1.2cm}
\caption
{
For more than two decades in time scale, 
the mass distribution obtained from our simulation shows a power-law behaviour 
at the low-mass end followed by a sharp exponential cut-off at the large-mass end, similar to Press-Schechter mass function
 which 
is commonly observed for real astrophysical data and
cosmological N-body simulations. 
This selfsimilar distribution function indicate an
underlying scale-invariant dynamics. }
\label{scalingfig}
\end{figure}

Topological scaling is believed to emerge from a self-similar growth process. 
An appropriate way to analyse dynamical scaling is 
to study the mass distribution function
$n(m,t)$, that is the fraction of aggregates of mass $m$ at time $t$.
The results of the measures of $n(m,t)$ at different times $t$ 
are shown in fig.(\ref{scalingfig}) as a function of $m/t$.
We have found, by observing various scaling 
behaviours of our model \cite{rl01}, that the mass distribution function has 
the self-similar form
\be
t^2 \cdot n(m,t)=\left({m\over t}\right)^3 e^{-(m/t)^2}
\label{scalingfinal}
\ee
as is shown in figure \ref{scalingfig}.

The $t^2$ factor is
a direct consequence of the mass conservation requirement of our model.

We have also analysed the density-density correlation 
for our model which develops a power law 
with exponent $-1$, followed by a flat behaviour.
In 1 dimension an exponent $-1$ implies single objects (or
dimension 0) followed by a smooth distribution.
In this respect, the power-law behaviour 
of the correlation 
does not signal a fractal measure. 
In this model, unlike in fractal structures, 
large and small scales for mass and void
distribution do not coexist, since the large mass particles are formed
by the destruction of the substructures. Hence, the increase in 
correlation length is basically determined by the separations of the nearest 
neighbours and since these distances grow linearly with time,  so
does the correlation length.

We mainly focus on the universal
 features of the distribution of the masses of
the aggregates and compare the self-similar 
features of such a function with that of Press-Schechter
frequently encounters in astrophysics \cite{ps74}.
Press-Schechter mass function is a manifestation of the self-similar nature of
gravitational hierarchical structure formation and has shown impressive 
agreements with simulations and observational results \cite{ps74}. 
Although the cellular automata presented here gives rise 
to a mass function similar in some respect to Press-Schechter, it does not 
have  the same exponents for white noise initial conditions.  
As compared to Press-Schechter distribution, here, small masses deplete 
rather fast and while they die out more massive objects do not form 
fast enough for the mass function to have a 
power-law decay for the small masses.

\section*{Acknowledgments}
R. M. is supported by TMR network ``Fractal structures
and Self-organization'' under the contract FMRXCT980183.
M. M. is supported by INFM through the project FORUM {\it 
clustering}

\smallskip
\smallskip



\end{document}